# Near-perihelion activity and composition of 3I/ATLAS from JUICE/MAJIS observations

D. Bockelée-Morvan[1], F. Poulet[2,5], Y. Langevin[2], B. Seignovert[3], C. Leyrat[1], G. Piccioni[4], C. Royer[2], S. Rodriguez[5], E. d'Aversa[4], R. Brunetto[2], J. Carter[2], T. Cavalié[6], M. De Sanctis[4], E. Lellouch[1], A. Migliorini[7], C. Pilorget[2], E. Quirico[8], S. Robert[9], F. Tosi[4]

[1]LIRA, Observatoire de Paris, Université PSL, CNRS, Sorbonne Université, Université de Paris, 5 place Jules Janssen, 92195 Meudon Cedex, France

[2]Institut d'Astrophysique Spatiale, CNRS/Université Paris-Saclay, 91405 Orsay Cedex, France

[3] Observatoire des Sciences de l'Univers Nantes Atlantique (Osuna/UARO), CNRS UAR-3281, Nantes Université, Nantes, France

[4]Istituto Nazionale di Astrofisica – Istituto di Astrofisica e Planetologia Spaziali (INAF-IAPS), Via del Fosso del Cavaliere 100, I-00133, Roma, Italy

[5] Université Paris Cité, Institut de physique du globe de Paris (IPGP), CNRS, Paris, France

[6]LAB, CNRS/Université de Bordeaux, Bordeaux, France

[7]INAF - Astronomical Observatory of Padova, Vicolo dell'Osservatorio, 5, 35122 Padova PD, Italy

[8]Université Grenoble Alpes, CNRS, IPAG, 38000 Grenoble, France

[9]Royal Belgian Institute for Space Aeronomy, 1180 Brussels, Belgium

**Abstract**

We present visible-to-infrared (0.5–5.56 µm) observations of the interstellar comet 3I/ATLAS obtained with the Moons and Jupiter Imaging Spectrometer (MAJIS) aboard the Jupiter Icy Moons Explorer (JUICE) spacecraft between 2025 November 2 and 25, shortly after perihelion. The fluorescence emission from $H_2O$ at 2.7 µm and $CO_2$ at 4.3 µm is detected at heliocentric distances of 1.36–1.68 au. A weak dust-scattered continuum is identified from which we determine the spectral slope, with values ranging from ~15% per 100 nm in the 0.65 − 0.9 µm region to 1-3% per 100 nm from 0.9 to 2.6 µm. Spatially resolved measurements show that the radial distributions of $H_2O$ and $CO_2$ species are consistent with release in the near-nucleus environment. We derive $H_2O$ production rates that decreased from $8 \times 10^{28}$ s$^{-1}$ to $4 \times 10^{28}$ s$^{-1}$ over the period November 2-25, while the $CO_2/H_2O$ ratio remained nearly constant at ~10%. The heliocentric evolution of the $CO_2$ production rate indicates activity controlled by solar heating. Combined with published CO measurements and the low gas expansion velocities, our analyses support a scenario in which $CO_2$ plays a major role in driving the activity of 3I/ATLAS near perihelion. In addition, broad emission features are identified in the 3.2–3.6 μm region that cannot be explained by the fluorescence of common cometary CH–bearing volatiles. Their spectral characteristics are consistent with aliphatic C–H functional groups and provide tentative evidence for the release of complex organic material from dust grains in the coma.

## 1. Introduction

Interstellar objects (ISOs) ejected during the dynamical evolution of planetary systems carry with them a record of the physical and chemical conditions that prevailed in their parent planetary disk. Therefore, the recent detections of ISOs during their passage through the solar system provides an unprecedented opportunity to investigate the nature of interstellar material originating from a distant planetary system. The first two confirmed ISOs, 1I/'Oumuamua and 2I/Borisov, revealed different physical properties: the former showed only indirect evidence of cometary activity (Oumuamua ISSI Team et al. 2019), while the latter displayed a volatile-rich coma with a composition remarkably similar to solar system comets but with an enhanced CO abundance (M. A. Cordiner et al. 2020; D. Bodewits et al. 2020). These initial detections have confirmed ISOs as a new class of small bodies, whose characterization is critical for constraining the diversity of planet formation across the Galaxy.

On July 1, 2025, the discovery of comet 3I/ATLAS on a hyperbolic trajectory by the Asteroid Terrestrial-impact Last Alert System (J. L. Tonry et al. 2018) provided a third opportunity to investigate the composition of an extrasolar icy body. Its large inbound velocity and dynamical properties suggest an origin in an old, low-metallicity stellar system, potentially associated with the Galactic thick disk (M. J. Hopkins et al. 2025; A. G. Taylor & D. Z. Seligman 2025). This potential source raises the possibility that its volatile inventory reflects a formation under conditions distinct from those of the solar system. Large anomalies in the D/H, $^{12}C/^{13}C$, and $^{14}N/^{15}N$ isotopic ratios were found, consistent with a comet material formed in a relatively cold and metal-poor environment early in the history of the Galaxy (M.A. Cordiner et al. 2026; C. Opitom et al. 2026; N.X. Roth et al. 2026c). Compared with comets of the solar system, 3I/ATLAS exhibited a low $HCN/H_2O$ production rate ratio near perihelion, an enhanced abundance of $CH_3OH$ relative to HCN, and an apparent depletion of sulfur relative to carbon (N. Biver et al. 2026; N.X. Roth et al. 2026a).

Despite extensive multi-wavelength observing campaigns (e.g., M. R. Combi et al. 2026; C.M. Lisse et al. 2026a, b), key questions remain regarding the spatial distribution and temporal evolution of its volatiles, particularly near-perihelion where nucleus outgassing is expected to be at its peak. Here, we present visible to near-infrared observations of 3I/ATLAS acquired by the Moons and Jupiter Imaging Spectrometer (MAJIS, F. Poulet et al. 2024) onboard the Jupiter Icy Moons Explorer (JUICE, O. Grasset et al. 2013*)* spacecraft during four observation windows: one obtained on 2025 November 2 only four days after perihelion (2025 October 30) and three acquired on 2025 November 12, 19, and 25 (Table A1 in Appendix A). These observations enable a time-resolved characterization of the inner coma of an interstellar comet at heliocentric distances ranging from 1.36 to 1.68 au. We report the detection of $H_2O$ emission (2.6-2.8 µm) and $CO_2$ emission (near 4.3 µm), as well as the detection of light scattered by dust. We also identify emission features in the 3.2-3.6 µm spectral range, which can be attributed to C-H bearing compounds. The fluorescence emissions of the $H_2O$ and $CO_2$ volatiles are analyzed through spectral and spatial retrievals and coma modeling, allowing to derive production rates and, by combining the MAJIS dataset with other observations, investigate the relative contributions of different outgassing drivers as a function of heliocentric distance.

Section 2 describes the MAJIS spectroscopic observations of 3I/ATLAS in the 0.5-5.55 µm spectral range. Section 3 focuses on the analysis of the rovibrational fluorescence emissions of $H_2O$ and $CO_2$. The processes driving the activity of 3I/ATLAS, along with the dust colors measured in the visible and near-infrared ranges and the potential carriers of the observed 3.4 µm band, are discussed in Section 4, followed by concluding remarks.

## 2. Observations and data reduction

The observation campaign of 3I/ATLAS was conducted during the hot-cruise phase of the JUICE mission, which corresponds to a JUICE-Sun distance smaller than 1.34 au. During this phase, the spacecraft had to maintain its high-gain antenna oriented toward the Sun to ensure thermal protection (F. Poulet et al. 2026). This configuration imposed significant operational constraints but still enabled a dedicated observing sequence spanning from 2025 November 2 to 25 UT (Figure 1). Within this observation window, data acquisition was limited to a small number of time slots due to thermal constraints. Yet, six comet-tracking sequences of 45 minutes each were successfully carried out, with the first one on November 2 at a heliocentric distance of ~1.36 au, only 4 days after perihelion. An additional four-hour extended sequence occurred on November 25 in a wait attitude, a fixed spacecraft-pointing mode in which the target was not actively tracked but instead drifted passively through the instrument field of view (FOV) due to relative motion. The campaign involved multiple instruments taking the advantage of the broad capabilities of the JUICE payload, originally designed for investigations of the icy satellites, Jupiter's atmosphere, and the Jovian magnetosphere (O. Grasset et al. 2013).

The spectral range of MAJIS is covered by two channels: the VISNIR (Visible and Near-Infrared) channel (0.5–2.35 µm) and the IR (Infrared) channel (2.28–5.56 µm). The spectral sampling is about 3.6 nm per band and 6.5 nm per band for the VISNIR and IR channels, respectively. The scientific FOV is 3.4° along the slit, centered on the optical boresight, and is sampled by 400 spatial elements with an Instantaneous Field Of View (IFOV) of 150 µrad (31 arcsec). The instrument can operate in both scan and pushbroom modes by moving or maintaining at fixed position the internal scan mirror, respectively. A single frame contains both spatial information along the sample axis (i.e., along the slit) of the VISNIR and IR detectors and spectral information along the wavelength axis. A hyperspectral data cube is built by repeating successive acquisitions in time along the second spatial axis (lines) by either the internal scan mirror or relative spacecraft pointing motion.

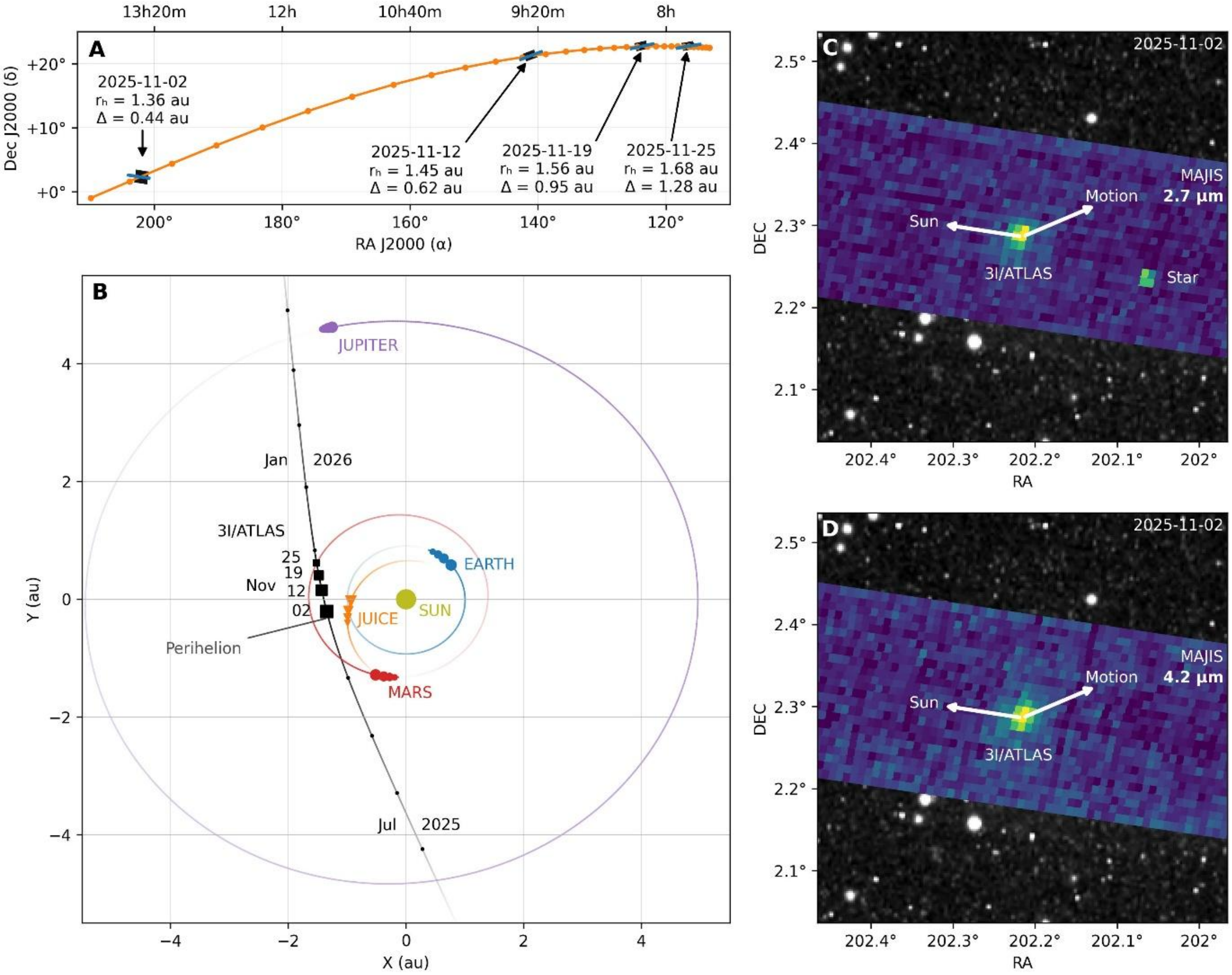

**Figure 1.** (A) The trajectory of 3I/ATLAS overlaid on the sky with the comet position at the dates of the MAJIS observations. The black squares show the NavCam field-of-view and the blue lines show the MAJIS slit. The heliocentric distance $r_h$ of the comet and the JUICE distance to the comet Δ are indicated. (B) Orbital motion of 3I/ATLAS (black square) projected onto the ecliptic plane. The positions of JUICE, Earth, Mars and Jupiter are shown for each observation slot. The retrograde orbit of 3I/ATLAS and its perihelion are marked. (C, D) Maps of $H_2O$ 2.7-µm emission (top panel), and $CO_2$ 4.3-µm emission (bottom panel) from 3I/ATLAS, extracted from a single cube (400 samples × 23 lines) acquired on 2025 November 2, at 06:03:38 UTC (see Section 3).

The MAJIS observations of 3I/ATLAS consisted of a 23-line scan performed with the scan mirror and centered on the predicted position of the comet. This corresponds to an angular coverage of approximately 0.2°, reflecting both pointing and temporal constraints. The data were acquired using a specific instrumental readout mode (100 Hz) optimized for low signal levels (Y. Langevin et al. 2022). Each line acquisition combined an integration time of 2 s with onboard frame averaging by a factor of four, resulting in an effective repetition time of 18.6 s and a duty cycle of 8 s. Reversible compression was selected (F. Poulet et al. 2024). During the observations, the instrument operated under stable thermal conditions, with the optical head (OH) maintained at ~133.3 K and the infrared focal plane array in the range of 87.5-89.2 K depending on the slot. In total, the campaign produced 14 hyperspectral science data cubes acquired on 2025 November 2, 12, 19, and 25 UT (Table A1).

MAJIS data were calibrated in spectral radiance using the instrument transfer function, which provides the conversion from digital numbers to radiance values under the assumption that the whole spectral response width of a MAJIS pixel is fulfilled by the incident light (Y. Langevin et al. 2024). The absolute calibration accuracy is on the order of 10% for both channels (Y. Langevin et al. submitted; S. Guerlet et al. submitted). Dark current and OH thermal contributions were removed by interpolating the signal from frames acquired immediately before and after each cube with the shutter closed. Noise levels of $10^{-5}$ and $5.10^{-6}$ in radiance (W/m²/sr/µm) per pixel are estimated for the VISNIR and IR channels, respectively. For signal extraction, we applied square-aperture photometry with sizes ranging from one MAJIS IFOV up to five IFOVs, depending on the signal strength of each spectral feature and science objectives. Minor pointing jitter occurred during spacecraft tracking, but did not affect the radiometric accuracy of the cometary signal. Residual background was further reduced by subtracting signals from lines and samples free of 3I/ATLAS signal. The observed flux was then mapped across the entire FOV of each cube to produce spatial maps for each cube (e.g., Figures 1 and B1) and to extract spectral features (Section 3). At the spatial resolution of the observations (~10,000 km per pixel at closest approach on 2025/11/2), detailed characterization of the dust and gas coma morphologies was not possible.

## 3. Results

### *3.1 Spectral signatures*

Figure 2 presents spectral analyses of 3I/ATLAS over the full spectral range of MAJIS data acquired on 2025 November 2. The scattering properties of the dust is expressed in relative reflectance (Figure 2A). Because the signal in each of the spectral elements of the VISNIR channel is very weak (< 30 e-), the reflectance shows a large dispersion. Therefore, the spectral slope was determined from a linear fit of the reflectance smoothed over 30 spectral elements. A red slope of 15 ± 3 % per 100 nm is measured between 0.6 and 0.9 µm. At longer wavelengths (0.9 to 1.5 µm), the spectral slope decreases significantly (2 ± 1 % per 100 nm). A weak continuum signal is also detected with the IR channel from 2.3 to 2.6 µm, with a reflectance consistent with the 2% per 100 nm slope observed from 0.9 to 1.5 µm, although there is a large uncertainty associated with this IR measurement.

Vibrational bands of $H_2O$, $CO_2$, and organic C-H-bearing molecules are detected in the spectral windows 2.6-2.8 µm, 4.2-4.3 µm, and 3.3-3.55 µm, respectively (Figures 2B-D). These emissions result from solar-pumped fluorescence of these molecules present in the coma (e.g., Bodewitts et al. 2024). $H_2O$ and $CO_2$ are detected at all observing dates, with band intensities decreasing by a factor 10-15 from November 2 to 25 (Table A1, Figure B1). The $H_2O$ and $CO_2$ comae are detected over several IFOVs (Figure B1), enabling the study of the radial variation of their emissions as a function of cometocentric distance in order to constrain their sources (Section 3.2). The 3.4-µm band is only detected on November 2. The robustness of this detection is supported by the map presented in Figure B2, which shows that the 3.4-µm excess is spatially coincident with the maxima of the $H_2O$ and $CO_2$ emissions.

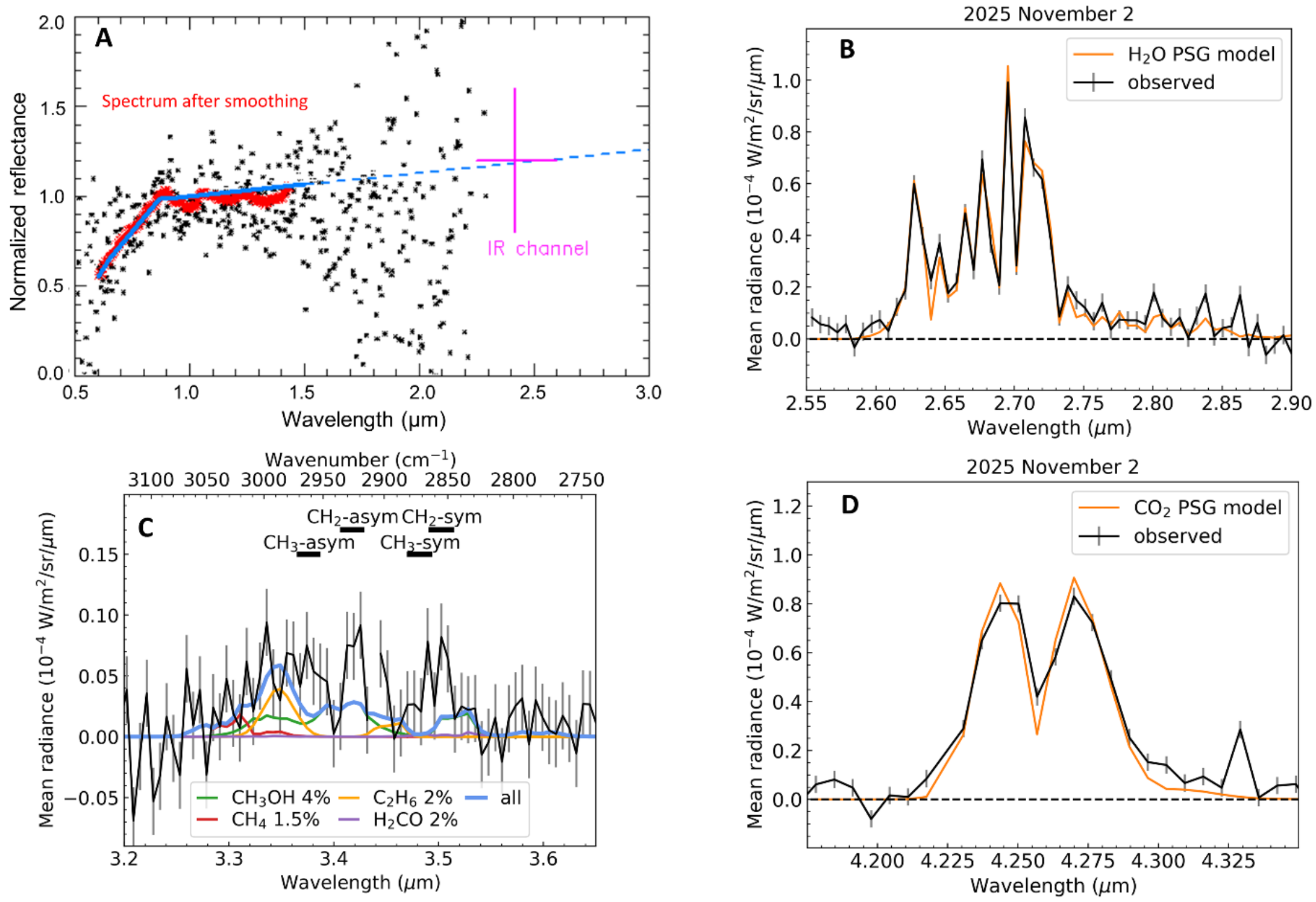

**Figure 2.** Sample spectra of 3I/ATLAS acquired on 2025 November 2 with MAJIS. Panel A: Normalized dust reflectance derived from the MAJIS measurements and scaled to unity at 1 µm. The black dots are the reflectance values evaluated for each spectral element of the VISNIR channel, while the red curve corresponds to the reflectance spectrum smoothed over 30 spectral elements. The two thick blue lines indicate linear fits over the 0.6-0.9 µm and 0.9-1.5 µm spectral ranges. The dashed blue line extends the 0.9-1.5 µm slope to 3 µm for comparison with the reflectance measured with the IR channel (magenta cross). Panels B, C, D: Major emission features from the mean of two central IFOVs, compared with synthetic spectra of fluorescence emission computed with the Planetary spectrum Generator (PSG, Sections 3.2, 4.3); B) $H_2O$ vibrational signature near 2.70 µm with best-fitted PSG spectrum (orange line, Section 3.2); C) 3.3-3.6 µm signatures attributed to C-H stretching modes. Synthetic PSG spectra of $CH_3OH$ (green), $CH_4$ (red), $C_2H_6$ (orange), and $H_2CO$ (purple), computed

assuming the abundances relative to $H_2O$ given in the inset, are overplotted together with their combined synthetic spectrum (blue); D) $CO_2$ band at 4.27 µm with best-fitted PSG spectrum (orange line).

*3.2. Radial and spectral modeling*

Figure 3 displays the radial profiles of the $H_2O$ and $CO_2$ band intensities with respect to the estimated photocenter on November 2 and 12. The dashed grey curves show the expected radial variation assuming isotropic release of these species from the nucleus at constant velocity (i.e., using the Haser model, L. Haser 1957). For these calculations, we adopted the expansion velocity $v_{\mathrm{exp}}$ of 0.6 km/s estimated from Doppler-resolved $H_2O$ spectra acquired from the JUICE/Submillimeter Wave Instrument (SWI) on November 2-6 (P. Hartogh, personal communication). We assumed photodissociation rates $\beta(H_2O)$ = 1.49 $10^{-5}$ $s^{-1}$ and $\beta(CO_2)$ = 3.2 $10^{-6}$ $s^{-1}$ at $r_h$ = 1 au, scaled as $r_h^{-2}$. These rates include corrections for solar activity relative to the values for quiet Sun reported by J. Crovisier (1989) for $H_2O$, and Huebner & Mukherjee (2015) for $CO_2$, based on the F10.7cm solar activity index. A FWHM equal to 1.47 times the IFOV size (31") for both the slit and pixel functions was assumed (G. Filacchione et al. 2024). Figure 3A also shows the expected $H_2O$ radial variation in the assumption of extended production of water with characteristic scale lengths of 1000 and 10,000 km.

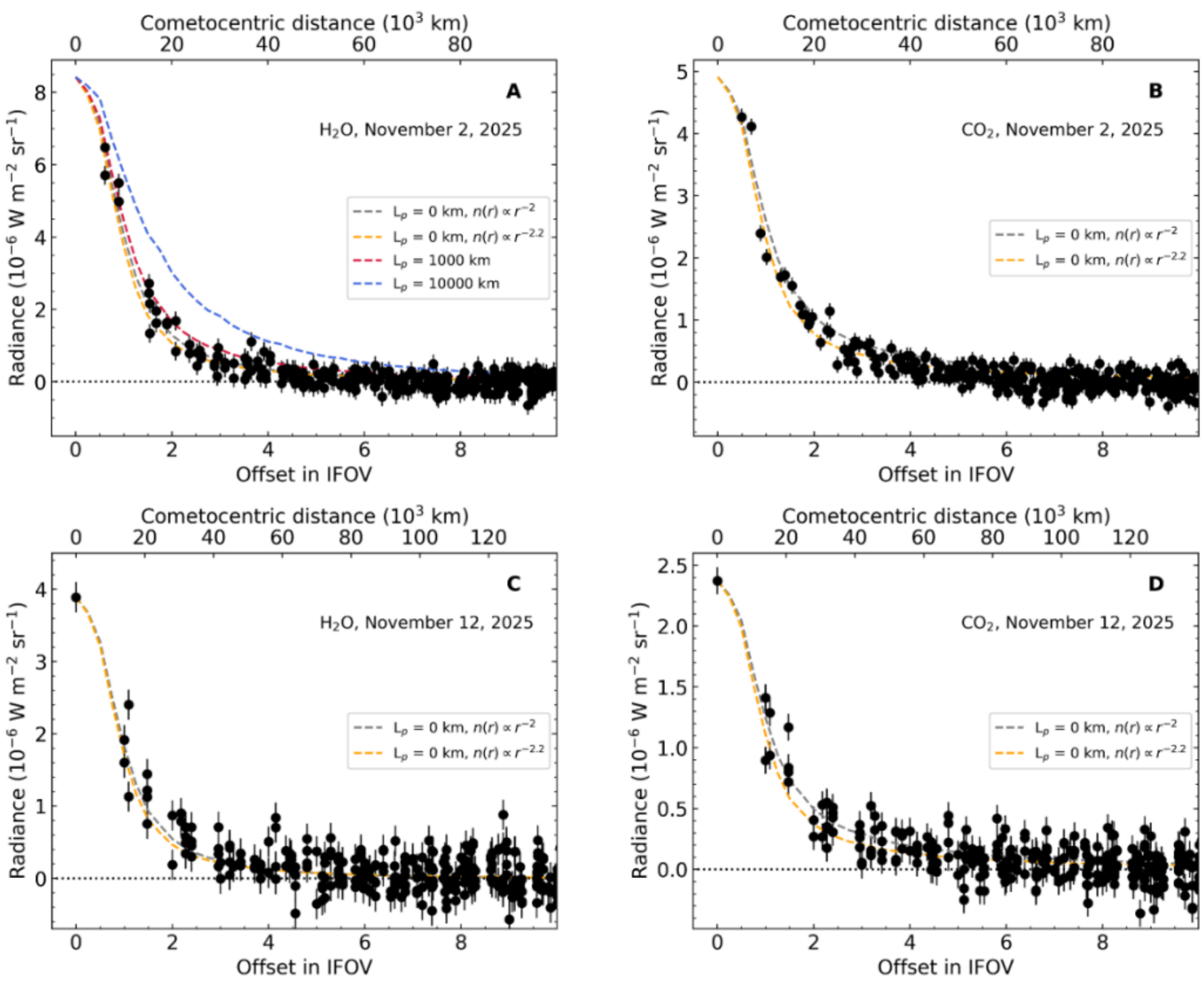

**Figure 3.** Radial distribution of the $H_2O$ (A, C) and $CO_2$ (B, D) band areas extracted from the maps obtained on 2025 November 2 and 12 (Figure B1). The dashed lines show the expected radial profiles for several descriptions of the local number density of $H_2O$ and $CO_2$ (see text and inserts). PSF smearing is considered.

The shapes of the observed $H_2O$ and $CO_2$ emission bands (Figures 2B and 2D) reflect the relative intensities of their constituent ro-vibrational lines, which are not individually spectrally resolved because of the limited spectral resolution of MAJIS ($\Delta\lambda$ ~ 6.3 nm). These relative intensities are directly related to the populations of the rotational levels in the ground vibrational state. We used the Planetary Spectrum Generator (PSG, Villanueva et al. 2018), which simulates fluorescence spectra of cometary molecules excited by solar infrared radiation and describes the level populations using a Boltzmann distribution characterized by a rotational temperature, $T_{rot}$. Figures 2B and 2D show PSG fits to the $H_2O$ and $CO_2$ spectra observed on November 2. The $H_2O$ band comprises emission from the $\nu_3$ and $\nu_1$ fundamental bands, as well as from combination bands (so-called hot bands) beyond 2.8 µm. $T_{rot}$ values derived from PSG fits to the spectra at the peak of $H_2O$ emission are listed in Table A2. The MAJIS observations sample $H_2O$ molecules that have left the collisional region, where rotational populations are thermalized to the gas kinetic temperature, and radiatively relax toward a cold fluorescence equilibrium. Because the local rotational temperature decreases with increasing cometocentric distance, the observations acquired on 12 and 19 November sampled more distant coma regions than those observed on 2 November. This naturally explains the lower effective rotational temperatures derived within the field of view. We used the non-LTE excitation model constrained by the observations of $CH_3OH$ millimeter lines in 3I/ATLAS on November 1-3 (N. Biver et al. 2026) to compute the expected $T_{rot}$ of water in the MAJIS extraction apertures. This non-LTE model computes the population of the rotational levels as a function of cometocentric distance, taking into account collisions with both neutrals and electrons, and radiative processes. Derived $T_{rot}$ values are in good agreement with MAJIS data (Table A2).

Inferred rotational temperatures from the $CO_2$ $\nu_3$ band at 4.27 µm are 80 ± 7 K, 87 ± 23 K, and 148 ± 20 K for November, 2, 12, and 19, respectively. Higher rotational temperatures for $CO_2$ wrt $H_2O$ is also observed at $r_h$ = 2.4 and 3.3 au in JWST observations of 3I/ATLAS (M.A. Cordiner et al. 2025, 2026; Roth et al. 2026b; Belyakov et al. 2026). The reported $CO_2$ rotational temperatures are lower than those derived from MAJIS near perihelion, except at 3.3 au pre-perihelion, where a value of ~90 K is obtained (M.A. Cordiner et al. 2025). The $CO_2$ band shape observed with MAJIS is not well reproduced by PSG (Figure 2D), in particular in the wings and the core of the band. This suggests the presence of $CO_2$ molecules at distinct rotational temperatures. Evidence for an additional hot $CO_2$ component is also provided by the JWST/NIRSpec data obtained at 3.3 au before perihelion and 2.4 au post-perihelion JWST/NIRSpec data, with the latter revealing high-energy $CO_2$ lines in the 4.2-4.4 µm spectral region. In contrast to $H_2O$, $CO_2$ lacks a permanent dipole moment, so pure rotational radiative transitions are forbidden. As the $CO_2$ molecules expand through the coma, their rotational temperature is expected to slowly evolve from the gas kinetic temperature prevailing in the collisional region, where collisions are sufficiently frequent to maintain thermal equilibrium, toward a hot fluorescence equilibrium temperature (J. Crovisier 1987). This behavior naturally explains the higher rotational temperatures measured for $CO_2$.

*3.3. Production rates*

Production rates were derived using the band intensities extracted in square boxes centered on the brightest IFOVs (Table A1). Most signal extractions were done in 5×5 IFOVs boxes in order to mitigate biases caused by the poor spatial sampling of the coma, while limiting the noise introduced by IFOV summation. The data acquired on November 2 show residual dust continuum emission shortward 2.5 µm affecting by 5% the $H_2O$ band intensity, which was corrected accordingly. The band intensities were integrated over spectral ranges encompassing most of the $H_2O$ and $CO_2$ emission and compared with the corresponding band intensities derived from PSG simulations. The same gas outflow velocity of 0.6 km/s was assumed for all observing dates, although the comet was receding from the Sun. Indeed, considering the gas acceleration in the coma (Section 4.2), we assumed that the expected decrease in the expansion velocity with increasing heliocentric distance would be at least partially compensated by the increasing distance between the comet and the JUICE spacecraft. The validity of this assumption cannot be tested, however, since no measurements of the heliocentric evolution of the gas velocity are available for 3I/ATLAS.

Table 1 lists the $H_2O$ and $CO_2$ production rates derived for the four observing dates using the photodissociation rates given in Section 3.2**.** Unlike $CO_2$, inferred $H_2O$ production rates depend slightly on the assumed $\beta(H_2O)$: adopting a value 20% higher increases the derived values by 5% (Nov. 2) to 10% (Nov. 19). We emphasize that the $H_2O$ detection is marginal for the last date, with a hint of detection only in the second over the eight observations conducted on this date. A possible explanation is that the peak of $H_2O$ coma coincided with the center of an IFOV only on that date, thereby enhancing its detectability. On November 25, $CO_2$ is also detected in only three of the eight cubes. The $H_2O$ production rate inferred on November 2 is consistent with that derived from contemporaneous $H_2O$ observations with JUICE/SWI (P. Hartogh, personal communication).

**Table 1**

$H_2O$ and $CO_2$ production rates

| Date | $r_h$ (au) | $Q(H_2O)$[a] ($10^{28}$ s$^{-1}$) | $Q(CO_2)$[a] ($10^{27}$ s$^{-1}$) | $Q(CO_2)/Q(H_2O)$[b] |
|---|---|---|---|---|
| 2025/11/ 2.25 | 1.363 | 8.0 ± 0.8 | 8.2 ± 0.8 | 0.103 ± 0.006 |
| 2025/11/12.30 | 1.447 | 6.8 ± 0.7 | 6.9 ± 1.0 | 0.101 ± 0.011 |
| 2025/11/19.30 | 1.555 | 4.7 ± 0.7 | 5.9 ± 1.4 | 0.125 ± 0.031 |
| 2025/11/25.63 | 1.678 | 4.1 ± 1.2 | 4.0 ± 0.7 | 0.098 ± 0.038 |

[a]The uncertainty includes a 10% uncertainty in the absolute radiometric calibration. [b]The uncertainty includes a 2% uncertainty in the relative calibration.

## 4. Discussion

### *4.1. VISNIR spectral slope*

The dust reflectance spectrum observed with the MAJIS VISNIR channel at $r_h$ = 1.36 au is broadly consistent with ground-based measurements at $r_h$ = 1.8 to 4.4 au, pre- and post-perihelion (e.g., R. de la Fuente Marcos et al., 2025; W.B. Hoogendam et al. 2025; L. Lazzarin et al. 2026; K. Medler et al. 2026; C. Opitom et al. 2025; D.Z. Seligman et al. 2025; B. Yang et al. 2025). All datasets display a steep red spectral slope in the visible domain that significantly decreases toward longer wavelengths (Figure C1). Since spectral slopes were determined over different wavelength intervals, Figure C1 shows the measured values against the midpoint wavelength of each interval. The spectral slope measured by MAJIS between 0.6 and 0.9 µm (15 ± 3 % per 100 nm) is consistent with the value of 18 ± 4 % per 100 nm reported by C. Opitom et al. (2025) from pre-perihelion observations at 4.5 au (0.5-0.9 µm spectral range). However, it is significantly higher than values reported at $r_h$ > 1.8 au by B. Yang et al. (2025), W.B. Hoogendam et al. 2025 and K. Medler et al. (2026) over the 0.65-0.9 µm range (4.4-9.6 % per 100 nm). This may suggest that the optical scattering properties of 3I/ATLAS's dust coma evolved as the comet approached perihelion. A similar trend was observed for 67P/Churyumov-Gerasimenko and was interpreted as resulting from the progressive dehydration of dust particles (G. Filacchione et al. 2020). In the NIR range ($\lambda$ > 0.9 µm), slightly blue colors were reported at $r_h$ = 1.8 to 3.6 au (K. Medler et al. 2026), whereas a red spectral slope was measured at $r_h$ = 4.1-4.4 au (B. Yang et al. 2025). MAJIS data at 1.36 au suggest a slightly red slope extending to 2.4 µm (Figure 2A), although the associated MAJIS uncertainties are large owing to the low signal level. The red slope in the visible domain suggests that the scattering is dominated by submicron dust grains rich in reddish materials (e.g., organic compounds which typically exhibit a red spectral slope). However, many organic-rich materials remain red beyond 0.9 µm, and therefore the spectral break may be the result of a process other than composition alone, particularly if the scattering becomes dominated by larger grains and thus by geometric optics due to the presence of particles larger than a few microns.

*4.2. Comet activity near perihelion*

The MAJIS observations of $H_2O$ and $CO_2$ emissions provide key information on the processes driving the activity of this comet. Figure 4A shows the $H_2O$ and $CO_2$ production rates derived from MAJIS data, together with values (including for CO) found in the literature, as a function of heliocentric distance. MAJIS provides unique information on the $CO_2$ outgassing activity at perihelion, complemented by observations of the same 4.27 µm $CO_2$ band acquired with JWST/NIRSpec and SPHEREx data at 3.1 and 3.3 au pre-perihelion, and at 2.1 and 2.4 au post-perihelion (M.A. Cordiner et al. 2025, 2026; C. M. Lisse et al. 2026a, b). Results from the 15 µm $CO_2$ band observed with JWST/MIRI (M. Belyakov et al. 2026), which differ from NIRSPec results, are not included in this analysis. The origin of this discrepancy remains unclear and may reflect an incomplete understanding of $CO_2$ excitation. Water production rate measurements are more numerous, with most of them derived from observations of the $H_2O$ photodissociation products H and OH (see references in the caption of Figure 4).

A striking result is that the water production rates derived from MAJIS data (7-8 $10^{28}$ $s^{-1}$ for November 2-12, Table 1) are four times lower than the values derived from large-aperture Lyα data from SOHO/SWAN (light blue symbols in Figure 4A), as reported by M. R. Combi et al. (2026). H. Tan et al. (2026) report, from the same SOHO/SWAN dataset, $Q(H_2O)$ values about 6 times lower than those reported by M.R. Combi et al. for the November 13-25 period, but still not consistent with MAJIS, lying on the opposite side of the discrepancy. If confirmed as robust, the high $Q(H_2O)$ values reported by Combi et al. (2026) would be consistent with icy-grain water release in the outermost coma (M. Combi et al. 2026). On the other hand, Figures 3A and 3C show that the radial distribution of $H_2O$ observed by MAJIS is consistent with a model assuming direct release from the nucleus. One possible interpretation is that MAJIS, owing to its much smaller field of view, primarily sampled the near-nucleus $H_2O$ component, whereas SOHO/SWAN was sensitive to a more extended source of water producing the H atoms. The sublimation of micrometric pure water-ice particles, with typical lifetimes of a few days at $r_h$ = 1.3-1.7 au (E. Zubko et al. 2026), would be the source of the additional water production observed by SOHO/SWAN. We note however that no evidence for such ice particles is found in the MAJIS VISNIR dust continuum, whose spectral slopes are consistent with those of the ice-free coma of 67P/Churyumov-Gerasimenko at perihelion (D. Bockelée-Morvan et al. 2019; G. Rinaldi et al. 2016, 2018; Filacchione et al. 2020).

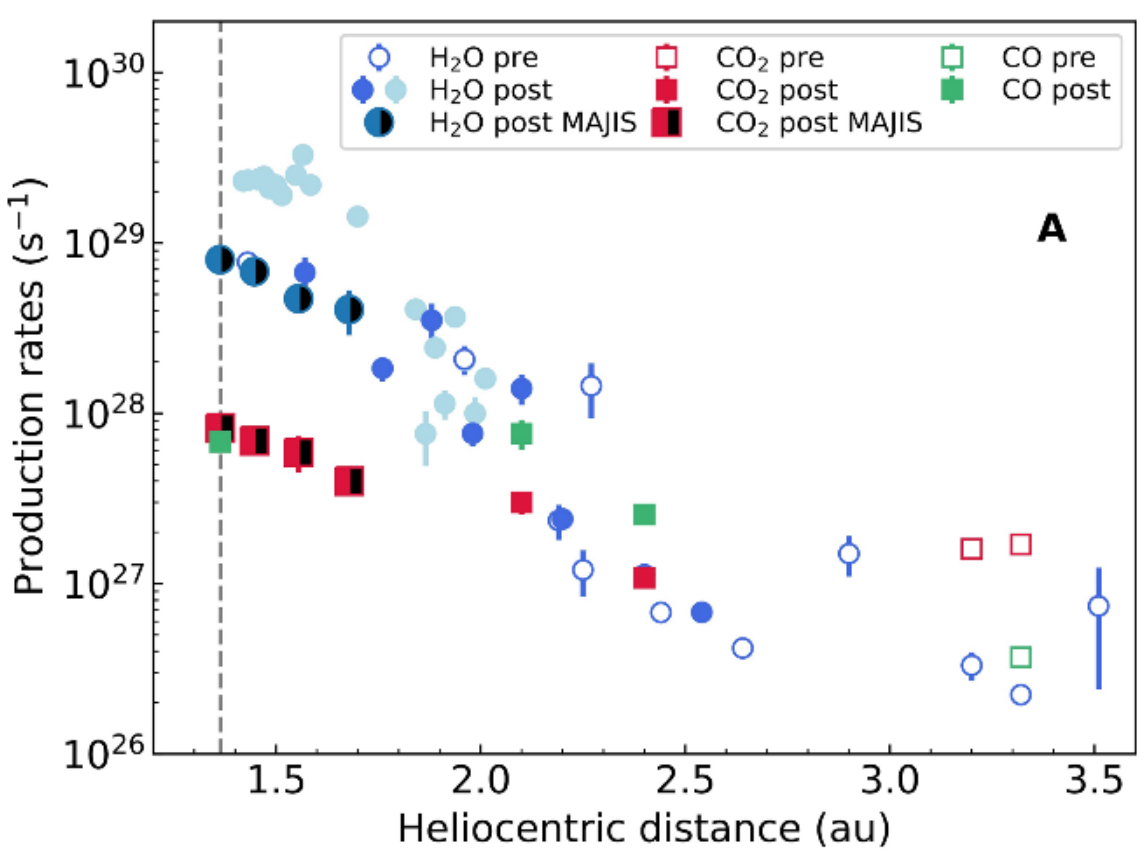

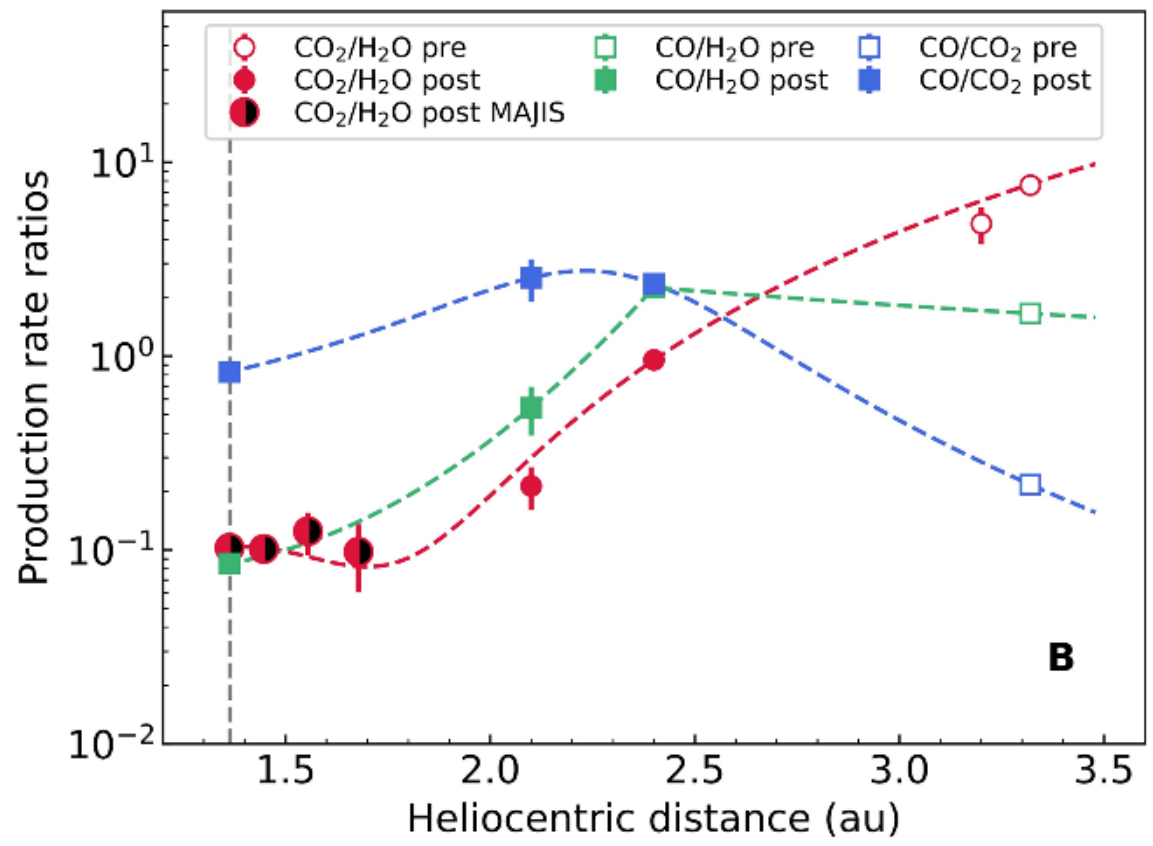

**Figure 4.** Heliocentric evolution of the $H_2O$, $CO_2$ and CO production rates (A) and their relative abundances (B). Panel A: $H_2O$ ($CO_2$) production rates are indicated by blue (red) symbols, with open and filled symbols for pre-perihelion and post-perihelion dates, respectively; green symbols are used for the CO production rates. Panel B: $CO_2/H_2O$, $CO/H_2O$ and $CO/CO_2$ ratios are indicated by red, green, and blue symbols, respectively; curves overplotted with dashed lines show the heliocentric trends. Data points using solely MAJIS data are plotted with bicolor symbols. Data other than from this work are from N. Biver et al. 2026 (CO, IRAM 30-m), M. A. Cordiner et al. 2025, 2026 ($H_2O$, $CO_2$, CO, JWST/NIRSpec), N.X. Roth et al. 2026 ($H_2O$, $CO_2$, CO, JWST/NIRSPec), M. Belyakov et al. 2026 ($H_2O$, JWST/MIRI, corrected using $v_{exp}$ = 0.3 km/s from M. A. Cordiner et al. 2026), M. R. Combi et al. 2026 (HI, SOHO/SWAN, shown in light blue), C. M. Lisse et al. 2026a, b ($H_2O$, $CO_2$, CO, SPHEREx), Z. Xing et al. 2025 (OH, UV, Neil Gehrels Swift Observatory), J. Crovisier et al. 2025 (personal communication, OH-18cm), J. Li et al. 2026 (OH-18cm) E. Jehin et al . 2025 (OH, UV, TRAPPIST), D. Hutsemékers et al. 2025 (OH, UV, Neil Gehrels Swift Observatory, VLT).

While long-lived pure-ice particles would contribute only marginally to the $H_2O$ production rate measured by MAJIS, the detected $H_2O$ molecules most likely originated from the sublimation of short-lived dirty icy grains near the nucleus, with a characteristic parent scale length constrained by the radial profile to $L_p$ << 1000 km (see red curve in Figure 3A). Indeed, the water production rate exceeds what would be expected if 100% of the surface was covered by water ice. Considering the 2.6 km nucleus diameter of 3I/ATLAS (M.-T. Hui et al. 2026) and a water production rate of $10^{29}$ s$^{-1}$, the active fraction of the nucleus, inferred following the approach of D. C. Lis et al. (2019), is on the order of 200%. This value is similar to the active fraction measured for 103P/Hartley (D. C. Lis et al. 2019), the prototype of hyperactive comets studied in detail by the Deep Impact spacecraft (M. F. A'Hearn et al. 2011). Hence, 3I/ATLAS possibly shares properties with comet 103P/Hartley 2, whose hyperactivity at perihelion was initially proposed to result from the direct sublimation of $CO_2$-rich regions of the nucleus, releasing icy grains and chunks in the coma (N. Fougere et al. 2013).

The $CO_2$ radial distribution observed on November 2 decreases more steeply than expected for isotropic outgassing from the nucleus at constant velocity (grey dashed line in Figure 3B), consistent with SPHEREx observations (C. M. Lisse et al. 2026b). A good reproduction of this $CO_2$ radial profile, as well as that observed on November 12, is obtained with a model in which the $CO_2$ local density is varying in $r^{-2.2}$, instead of the Haser $r^{-2}$ dependence with cometocentric distance $r$ (orange dashed curves in Figures 3B, C). A $\chi$-squared analysis yields a 3-$\sigma$ uncertainty of 0.1 on the power-law index. Simulations with a $r^{-2.2}$ dependence of the $H_2O$ number density also explain the observed $H_2O$ radial profiles (Figures 3A, C). Gas acceleration can lead to this behavior, and would imply a ~25% increase in the gas expansion velocity between 10,000 and 40,000 km. Such acceleration at these distances is expected as the result of photolytic heating (M. R. Combi et al. 1999). Additional evidence for gas acceleration in the coma of 3I/ATLAS comes from the significantly lower gas velocities of 0.4 km/s deduced from observations sampling inner coma regions (N. Biver et al. 2026), compared with the value of ~0.6 km/s measured by SWI at the same dates with a larger field of view.

The heliocentric evolution of the $CO_2$ production rate follows roughly $r_h^{-1.9}$ and $r_h^{-3.7}$ trends, for pre-perihelion (3.3-1.37 au) and post-perihelion (1.36-2.4 au), respectively, showing that the $CO_2$ activity responded to the solar flux received at the nucleus surface and suggesting that $CO_2$ was released from the surface or shallow subsurface layers (e.g., A. Guilbert-Lepoutre et al. 2016). The pre-/post-perihelion activity asymmetry observed for $CO_2$ is also observed for the dust, though with significantly steeper power-law exponents (M.-T. Hui et al. 2026). Water also displays a steep heliocentric evolution.

The $CO_2/H_2O$ production rate ratio measured in 3I/ATLAS at $r_h$ = 1.36-1.68 au is ~ 10% (Table 1), consistent with values measured in solar system comets at similar heliocentric distances (O. Harrington Pinto et al. 2022). The increase in the $CO_2/H_2O$ ratio with increasing $r_h$ (reaching ~800% at $r_h$ = 3.3 au pre-perihelion; Figure 4B) is naturally explained by the increasing sublimation lifetimes of icy particles causing the fast decrease of the $H_2O$ production. Moreover, the $CO_2/H_2O$ ratio measured near perihelion is most likely a lower limit to the production rate ratio at the nucleus surface, considering the contribution of short-lived icy grains to the water production rate measured by MAJIS. A scenario in which the surface activity was driven by $CO_2$ sublimation is supported by the unusually low expansion velocities of coma gases measured in 3I/ATLAS, which can be explained by near-nucleus gas flow driven by heavy molecules (N. Biver et al. 2026; M.A. Cordiner et al. 2026). In contrast, a recent study of Deep Impact observations of comet 103P/Hartley 2 argues for significant extended production of $CO_2$ from grains and instead suggests that its activity was primarily driven by $H_2O$ sublimation (Y. Shou et al. 2025). This is consistent with the expansion velocities of 103P coma gases which are typical of those commonly observed in comets with water-driven activity (N. Biver et al. 2011; J. Boissier et al. 2014).

CO production rate measurements are available for November 1-3 (N. Biver et al. 2026), leading to a $CO/CO_2$ production rate ratio of ~1 for comet 3I/ATLAS near perihelion. Comets display a wide range of $CO/CO_2$ abundances, spanning values from 0.01 to 10 (O. Harrington Pinto et al. 2022) for those observed within 2 au from the Sun, with the lowest value measured for 103P/Hartley 2. As shown in Figure 4B, the $CO/CO_2$ production-rate ratio varied significantly with heliocentric distance, with $CO_2$ dominating over CO at large pre-perihelion $r_h$, and CO dominating over $CO_2$ at $r_h$ = 2.1-2.4 au post-perihelion. Post-perihelion images of the CO and $CO_2$ comae obtained with JWST reveal highly asymmetric and distinct outgassing patterns for the two species, emphasizing the complexity of the activity of 3I/ATLAS (N.X. Roth et al. 2026c).

*4.3. Evidence for complex organic emission*

Cometary spectra obtained in the 3.2-3.6 µm region at high ($R$ > 20,000) or moderately high ($R$ = 3000) spectral resolution show vibrational fluorescence emissions from $CH_4$, $C_2H_6$, $CH_3OH$ and $H_2CO$ (N. Biver et al. 2024, and references therein). Figure 2C shows the expected signatures of these molecules, computed with PSG, superimposed on the MAJIS spectrum of 3I/ATLAS obtained on November 2. We assumed the $CH_3OH/H_2O$ relative abundance measured in 3I/ATLAS at IRAM (N. Biver et al. 2026), and $CH_4/H_2O$ and $C_2H_6/H_2O$ abundances of 1.5% and 2%, respectively, in the upper range of values measured in comets (N. Biver et al. 2024). We set the $H_2CO/H_2O$ abundance ratio to an arbitrarily high value of 2%, i.e., five times higher than the value measured in 3I/ATLAS (N. Biver et al. 2026), in order to account for possible extended $H_2CO$ production. The MAJIS spectrum does not resemble the PSG synthetic spectrum, which is instead similar to typical cometary spectra (e.g., C. E. Woodward et al. 2025; N. Biver et al. 2024), including the spectrum of 3I/ATLAS observed at $r_h$ = 2.4 au (N.X. Roth et al. 2026c). Residual emission features are observed at wavelengths characteristic of the symmetric and antisymmetric stretching vibrational modes of $CH_2$ and $CH_3$ functional groups (Figure 2C), although the low S/N of this measurement should be kept in mind. The pronounced signatures span the spectral ranges 2845-2871 $cm^{-1}$, 2912-2936 $cm^{-1}$, and 2950-3020 $cm^{-1}$. A peak is observed at ~2900 $cm^{-1}$, consistent with the position of a Fermi resonance in alkyl chains or a C-H stretching vibration. No aromatic C-H signature is detected. Attempts to assign these features to simple molecules (e.g., $C_2H_5OH$, $CH_2CHCH_3$, $CH_3CH_2NH_2$) were unsuccessful. Their non-detection at $r_h$ = 2.4 au (N. X. Roth et al. 2026c) would then suggest that their precursors are released into the coma through a process that is strongly temperature-dependent (e.g., the degradation of organic material present in dust particles). Supporting this interpretation, the ROSINA mass spectrometer aboard the Rosetta mission identified a large variety of complex gas-phase molecules in the coma of 67P/Churyumov-Gerasimenko, thought to be thermally desorbed from dust particles, with chain-based hydrocarbons composing the most abundant group (N. Hänni et al. 2022). Emissivity signatures arising from thermal emission by organic-rich dust particles can also be considered. Nevertheless, the spectral characteristics produced by such materials depend on multiple structural parameters (R.A. MacPhail et al. 1984), making a comprehensive assessment beyond the scope of this paper. Since the dust continuum emission in this spectral region, expected to consist of both thermal emission and scattered light, is weak and remains undetected (Figure 2C), it is unclear whether such high band contrast can be explained by this mechanism.

## 5. Implications

The passage of 3I/ATLAS through the solar system offered a new opportunity to investigate the chemical composition of an icy planetesimal formed in another planetary system. Unlike 2I/Borisov, whose composition was found to be remarkably similar to that of solar system comets except for an enrichment in CO, 3I/ATLAS exhibited several characteristics that distinguish it from solar system comets (N. Biver et al. 2026, M.A. Cordiner et al. 2026, C. Opitom et al. 2026, N.X. Roth et al. 2026a-c). Evidence that 3I/ATLAS formed in a chemically distinct extrasolar environment also comes from the $CH_2$ and $CH_3$ stretching signatures observed in the MAJIS spectra, which suggest a high abundance of hydrocarbons in the dust grains, although the low signal-to-noise ratio warrants caution.

MAJIS observations of $H_2O$ and $CO_2$, complemented by radio observations, show that 3I/ATLAS' coma presented a $CO_2$-rich and CO-rich composition at perihelion. The C/O ratio inferred from the perihelion abundances of CO, $CO_2$, $H_2O$ and $CH_3OH$ is $0.17 \pm 0.01$, placing it in the upper range of values measured in solar system comets at similar heliocentric distances (O. Harrington Pinto et al. 2022; D.Z. Seligman et al. 2022). C/O values in the range 0.4-0.6 are inferred using data at $r_h > 2$ au. Such high C/O ratios are consistent with formation in the outermost regions of the natal protoplanetary disk, beyond the $CO_2$ snow line (D. Z. Seligman et al. 2022). A similar conclusion was reached for 2I/Borisov (D.Z. Seligman et al. 2022).

With an active fractional area of 200% at perihelion, as measured from MAJIS data, 3I/ATLAS belongs to the class of hyperactive comets that release icy grains and chunks into the coma. Using a sample of comets with determined nucleus size and water production rates near perihelion, D.C. Lis et al. (2019) showed that all comets with effective nucleus radii smaller than 1.2 km are hyperactive. Hence, given its nucleus radius of 1.3 km, the hyperactivity of 3I/ATLAS is not exceptional. 2I/Borisov, whose nucleus radius is below 0.4 km (M.-T. Hui et al. 2020), was also hyperactive, as suggested by the active fractional area estimated to > 90% at 2.38 au, compared with only 15% for 3I/ATLAS at the same heliocentric distance. This suggests that the large amount of subliming icy aggregates in hyperactive comets, including 2I/Borisov and 3I/ATLAS, is possibly not related to a higher ice-to-refractory content, but rather to their size. 3I/ATLAS nevertheless remains exceptional among well-studied hyperactive comets. Unlike 103P/Hartley 2, whose activity was primarily water-driven according to a recent study, its near-perihelion activity appears to have been dominated by the sublimation of the heavy volatiles $CO_2$ and CO.

## Acknowledgments

The authors acknowledge financial support from the Centre national d'études spatiales (CNES), France (ROR: https://ror.org/04h1h0y33), within the framework of the JUICE space mission. This work has been developed under the ASI-INAF agreement no. 2023-6-HH.0. The authors also wish to thank ESA teams from SOC and MOC for their technical and operational support to the MAJIS project. JUICE is a mission under ESA leadership with contributions from its Member States, NASA, JAXA and the Israel Space Agency. It is the first Large-class mission in ESA’s Cosmic Vision programme. D.B.M. and F.P. acknowledges useful discussions with O. Poch and G. Filacchione.

## Appendix A: $H_2O$ and $CO_2$ band intensities measured in 3I/ATLAS with MAJIS

**Table A1**

$H_2O$ and $CO_2$ band intensities measured in 3I/ATLAS with MAJIS

| Date | # observations | $r_h$ (au) | Δ (au) | $H_2O$ FOV[a] (IFOV) | $I_{H2O}$[b,c,d] (W $m^{-2}$ $sr^{-1}$) | $CO_2$ FOV[a] (IFOV) | $I_{CO2}$[e,c,d] (W $m^{-2}$ $sr^{-1}$) |
|---|---|---|---|---|---|---|---|
| 2025/11/ 2.25 | 2 | 1.363 | 0.444 | 4×4 | $2.65 \pm 0.06\ 10^{-6}$ | 5×5 | $1.24 \pm 0.03\ 10^{-6}$ |
| 2025/11/12.3 | 2 | 1.447 | 0.621 | 5×5 | $9.57 \pm 0.42\ 10^{-7}$ | 5×5 | $6.45 \pm 0.22\ 10^{-7}$ |
| 2025/11/19.3 | 2 | 1.555 | 0.950 | 5×5 | $3.48 \pm 0.40\ 10^{-7}$ | 5×5 | $2.95 \pm 0.21\ 10^{-7}$ |
| 2025/11/25.63 | 8 | 1.678 | 1.274 | 3×3 | $3.67 \pm 1.00\ 10^{-7}$ | 5×5 | $1.20 \pm 0.16\ 10^{-7}$ |

a) Dimensions of the square box, in IFOVs, from which the signal is extracted. b) $H_2O$ band area computed over 2.621-2.732 µm. c) The uncertainty of 10% in the radiometric calibration is not included. d) For November 2, 12, and 19, average of the two acquired hyperspectral cubes; for $CO_2$ on November 25, average of three cubes (second, third and fifth, over the eight acquired cubes); for $H_2O$ on November 25, only the second cube showing hint of $H_2O$ emission is considered. e) $CO_2$ band area computed over 4.217-4.303 µm.

**Table A2**

$H_2O$ rotational temperatures on the brightest IFOVs

| Date | Measured $T_{rot}$ (K) | Model $T_{rot}$ (K) |
|---|---|---|
| 2025/11/ 2.25 | 46 ± 4 | 43 |
| 2025/11/12.3 | 36 ± 6 | 38 |
| 2025/11/19.3 | 31 ± 8 | 33 |

## Appendix B: Maps of the $H_2O$ 2.7-µm, $CO_2$ 4.3-µm, and 3.4-µm bands

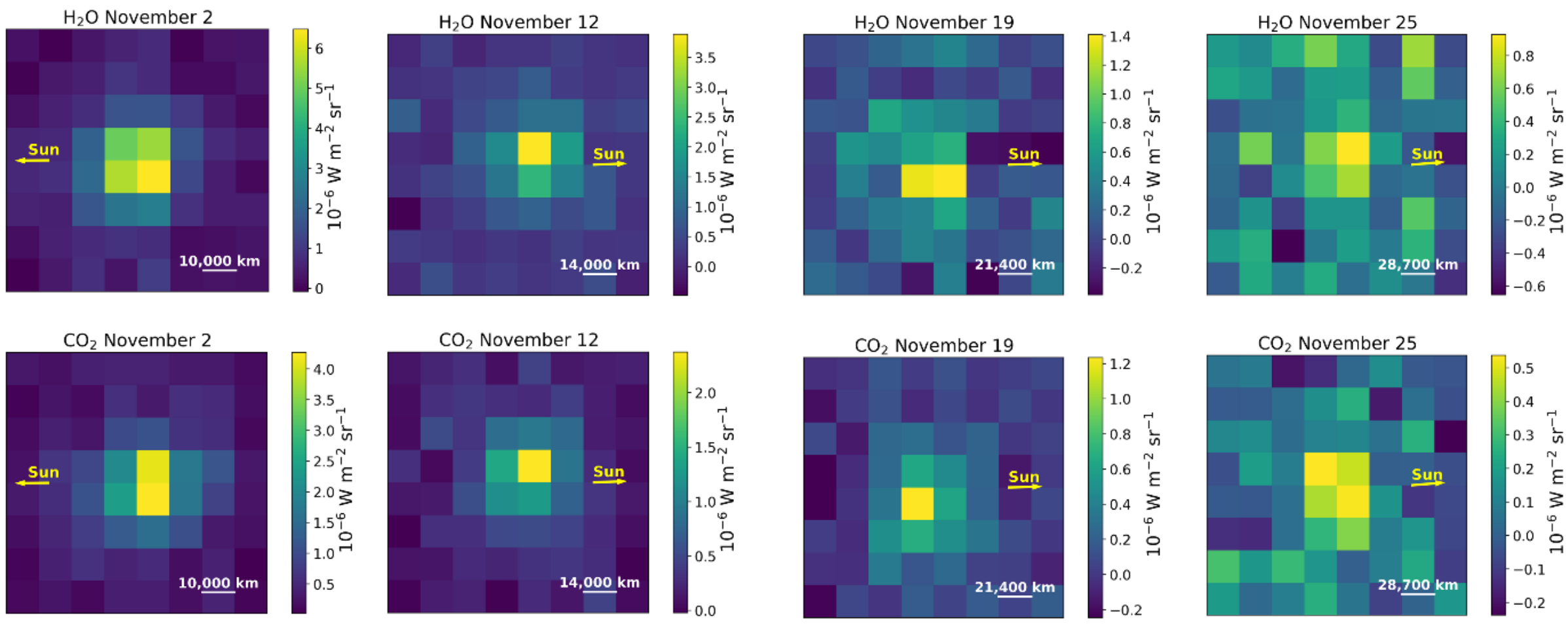

**Figure B1.** Maps of the $H_2O$-2.7 µm band (top) and $CO_2$-4.27 µm band (bottom) observed on 2025 November 2, 12, 19 and 25 (from left to right). The band areas (in W $m^{-2}$ $sr^{-1}$) have been computed over 2.621-2.732 µm, and 4.217-4.303 µm, for $H_2O$ and $CO_2$, respectively.

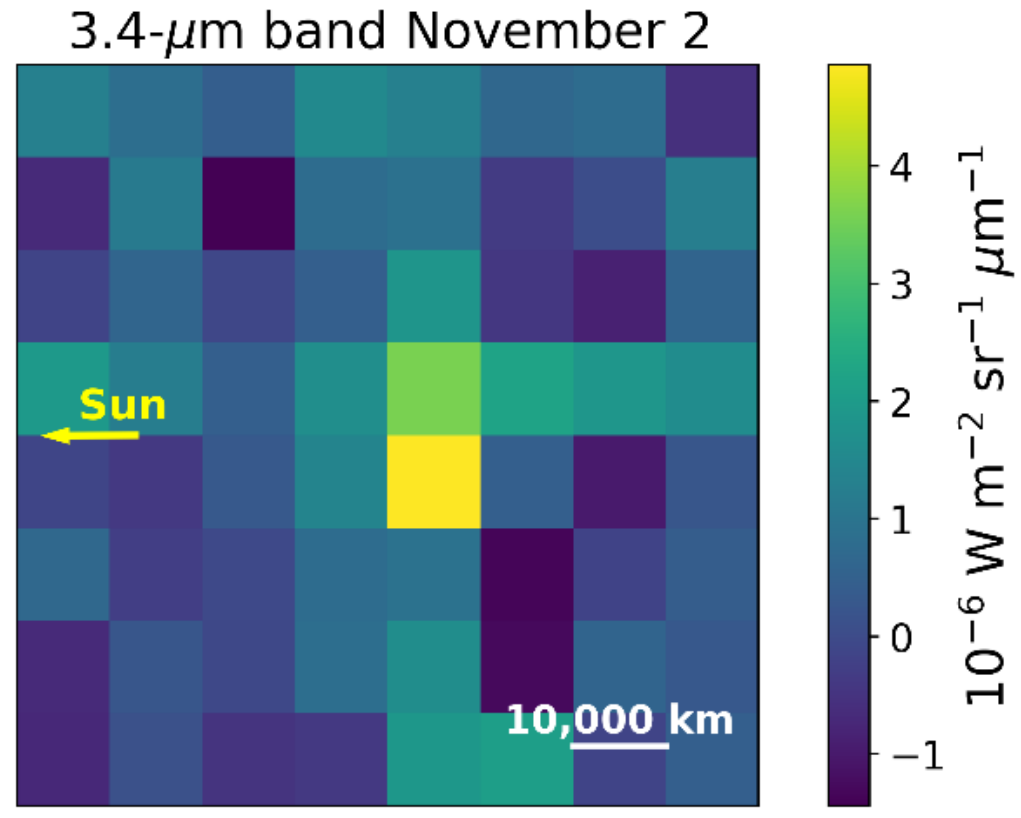

**Figure B2.** Map of the 3.4-µm band detected in 3I/ATLAS on 2025 November 2. The intensity is the median radiance of the spectral channels spanning the 3.30-3.55 µm wavelength interval. The FOV is identical to that shown for November 2 in Figure B1.

**Appendix C: Dust scattering properties in 3I/ATLAS**

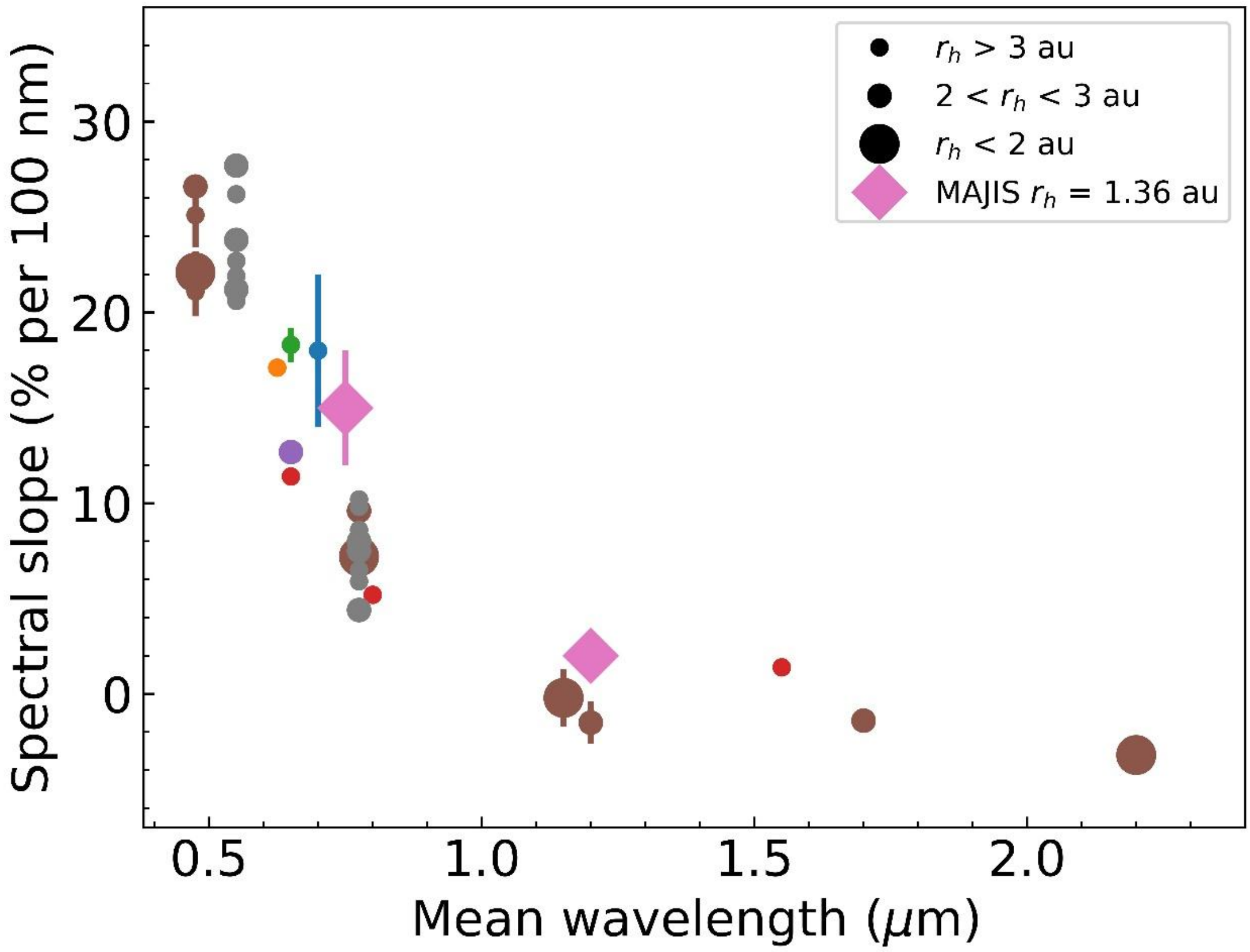

**Figure C1**. Spectral slopes from reflectance spectra of the dust coma of 3I/ATLAS. The MAJIS measurements are shown as pink diamonds while dots correspond to data from R. de la Fuente Marcos et al. 2025 (green dots), W.B. Hoogendam et al. 2025 (gray), L. Lazzarin et al. 2026 (purple), K. Medler et al. 2026 (brown), C. Opitom et al. 2025 (blue), D.Z. Seligman et al. 2025 (orange), and B. Yang et al. 2025 (red). Spectral slopes were determined over different wavelength ranges and plotted against the midpoint wavelength of each interval. Different symbol sizes are used to indicate the heliocentric distance at which the data were acquired.